\documentclass[pre,aps,reprint,amsmath]{revtex4-1}

\usepackage[T1]{fontenc} 
\usepackage{graphicx}
\usepackage[utf8]{inputenc}
\usepackage{lmodern}
\usepackage{bm}
\usepackage{bbold}
\usepackage{hyperref}

\begin{document}
\title{Non-negative Interfacial Tension in Phase-Separated Active Brownian Particles}

\author{Sophie Hermann}
\affiliation{Theoretische Physik II, Physikalisches Institut, Universit{\"a}t Bayreuth, D-95447 Bayreuth, Germany}

\author{Daniel de las Heras}
\affiliation{Theoretische Physik II, Physikalisches Institut, Universit{\"a}t Bayreuth, D-95447 Bayreuth, Germany}

\author{Matthias Schmidt}
\affiliation{Theoretische Physik II, Physikalisches Institut, Universit{\"a}t Bayreuth, D-95447 Bayreuth, Germany}
\email{Matthias.Schmidt@uni-bayreuth.de}

\date{1 August 2019, Phys. Rev. Lett. \textbf{128}, 268002 (2019).}

\begin{abstract}
We present a microscopic theory for the nonequilibrium interfacial
tension $\gamma_{\rm gl}$ of the free interface between gas and liquid
phases of active Brownian particles.  The underlying square gradient
treatment and the splitting of the force balance in flow and
structural contributions is general and applies to inhomogeneous
nonequilibrium steady states.  We find $\gamma_{\rm gl}\geq 0$, which
opposes claims by Bialk\'e et al.\ [Phys. Rev. Lett.  {\bf 115},
  098301 (2015)] and delivers the theoretical justification for the
widely observed interfacial stability in active Brownian dynamics
many-body simulations.
\end{abstract}

\maketitle
The interfacial tension (or ``surface tension'') of the free interface between 
two coexisting bulk phases is one of the most important quantities in the description of a
wide range of interfacial phenomena. The tension $\gamma_{\rm gl}$
between coexisting gas and liquid bulk phases plays a particularly
central role due to the high symmetry of the coexisting fluid phases.
It is a key quantity in the Kelvin equation for capillary condensation, for the strength of the thermal capillary wave
spectrum, and for the Laplace pressure in droplets.

The typical values of the interfacial tension vary over many orders of
magnitude, when going from molecular to colloidal systems. Using the
particle size $\sigma$ and the thermal
energy $k_BT$ as the natural scales, the scaled interfacial tension $\gamma_{\rm gl}
\sigma^2/k_BT$ is typically of the order of unity. The dependence on
$\sigma$ is particularly dramatic when going from atoms 
to colloids. An associated factor of $10^3$ of increase in length
scale translates into a decrease of $\gamma_{\rm gl}$ by a factor of
$10^{-6}$, as e.g.\ theoretically
\cite{vrij1997,brader2000,brader2002} and experimentally
\cite{hoog1999,hoog2001,aarts2004} demonstrated in phase separated
colloid-polymer mixtures, where confocal microscopy can be used to
great effect in studying, e.g., droplet coalescence \cite{aarts2008} and
viscous fingering \cite{setu2013}.

Very notably, the existence of the interfacial tension is the
mechanism by which macroscopic fluid interfaces, such as in droplets
and soap bubbles, attain a minimal geometric shape. The phase 
separated system minimizes the product of $\gamma_{\rm gl}$ and the 
interfacial area of the interface. As $\gamma_{\rm gl}$ is independent of
curvature in a first approximation, this amounts to minimizing the
interfacial area alone. This effect is, e.g.,\ commonly exploited in
microscopic computer simulation work, where the use of periodic
boundary conditions and suitable elongated box geometries offers the
system a preferred (short) direction for the choice of interface
orientation, and hence a stabilizing mechanism that truncates large
scale fluctuations. 
This also applies to active Brownian particles, i.e.,\ colloids where the diffusive motion is supplemented by directed self-propulsion and which phase separate at large enough swimming strength \cite{farage2015,brader2017,speck2015,utrecht2018}. 
Typical experiments rely on catalyzing a chemical reaction to induce such ``swimming'' \cite{buttinoni2013}.

There is much current progress in the description of free equilibrium
interfaces, such as, e.g.,\ geometry-induced
capillary emptying \cite{parry2016}, the local structure factor near
an interface \cite{parry2016cwt} and Goldstone modes and resonances in
the fluid interfacial region \cite{parry2019}.
A variety of related deep theoretical topics have been addressed
recently, including the curvature dependence of the surface free
energy of liquid drops and bubbles \cite{oettel2010}, the adsorption
of nanoparticles at fluid interfaces \cite{oettel2007}, the free
energy of complex-shaped objects \cite{roth2004}, the characterization
of the ``intrinsic'' density profile for liquid surfaces
\cite{tarazona2001,tarazona2004,tarazona2005}, and the interface
tension of curved interfaces \cite{oettel2012}.

All of the above physical understanding is necessarily based on the
fundamental property $\gamma_{\rm gl}\geq 0$.  This seemingly
indisputable fact was recently challenged based on computer simulation
work by Bialk\'e et al.\ \cite{bialke2015} in active Brownian
particles. The authors of Ref. \cite{bialke2015} used the 
pressure tensor route and found their results for the interfacial
tension to be negative.  They argue that this ``is a genuine
nonequilibrium effect that is rationalized in terms of a positive
stiffness.''
Patch et al.\ \cite{patch2018} reproduce the negative result
using an expression for $\gamma_{\rm gl}$ similar to that of
Ref.~\cite{bialke2015}, but with a different method for calculating
the active contribution.
From analysis of the interfacial (capillary wave) fluctuations both
groups find a positive value for the interfacial stiffness
\cite{bialke2015,patch2018}.
Lee constructs a coarse-grained model with an effective surface
tension that is positive, and he is able to describe his simulation
data \cite{lee2017}.
Solon et al.~\cite{solon2018} in their numerical analysis find a
negative value for the tension, but they also state that their
framework supports both positive and negative values.
Marconi and Maggi~\cite{marconi2015} state that the tension would turn
out to be negative in their theory. Subsequently, Marconi et
al.~\cite{marconi2016} through analytical work have reconsidered the
problem of the mechanical derivation of $\gamma_{\rm gl}$, but these
authors do not report numerical results from their theory and they
do not comment on the sign of $\gamma_{\rm gl}$ in
Ref.~\cite{marconi2016}.
Das et al.~\cite{das2019} investigated different expressions for
the microscopic stress. The authors state that in their treatment
the surface tension of active systems can be determined, but they
have not done so in Ref.~\cite{das2019}.
Considering the influence of activity on the gas-liquid interface of
the Lennard-Jones system, Paliwal et
al.~\cite{paliwal2017acvtiveLJinterface} use the pressure tensor route
and find a negative contribution from their swim term, but overall
positive values for $\gamma_{\rm gl}$ across a wide parameter range.

Here we demonstrate, based on a nonequilibrium generalization of the
microscopic treatment of the interface pioneered by van der Waals
\cite{widom}, that indeed the tension $\gamma_{\rm gl} \geq 0$ for
phase-separated active Brownian particles. Its scaled value in natural
units is of order unity, and vanishes with a 3/2 (mean-field) exponent
near the critical point. This proves, on a sound theoretical footing,
the hitherto unexplained stability of the planar active gas-liquid
interface and demonstrates the route ahead to the quantitative
description of nonequilibrium interfacial properties and phenomena.
Our treatment is based on discriminating between structural forces
that generate the tension and the flow force balance which does not.

Our mechanism for bulk phase separation is based on the exact translational one-body force balance equation \cite{krinninger2019, krinninger2016}
\begin{align}
\gamma \textbf{v} = \textbf{f}_\text{id} + \textbf{f}_\text{int} + \gamma s \boldsymbol{\omega}, \label{eq:force}
\end{align}
where the friction force on the left-hand side is balanced by the ideal diffusive force $\textbf{f}_\text{id}$, the internal force $\textbf{f}_\text{int}$ and the free swim force $\gamma s \boldsymbol{\omega}$ on the right-hand side. The friction constant is indicated by $\gamma$ and $s$ denotes the constant free swim speed. The velocity $\textbf{v}$, the density $\rho$, $\textbf{f}_\text{id}$, and $\textbf{f}_\text{int}$ all depend on position $\textbf{r}$ and orientation $\boldsymbol{\omega}$, but not on time as we are considering steady states.
Furthermore, we assume the interface between the dense (liquid) and dilute (gas) phase as perpendicular to the $x$ axis and translational invariance with respect to other spatial coordinates. Hence the density varies along the $x$ axis of the system.
The ideal diffusive force field is given exactly as $\textbf{f}_\text{id} = - k_\text{B}T \nabla \ln \rho$. 
The internal force field consists of adiabatic and superadiabatic contributions and is defined as 
\begin{align}
\textbf{f}_\text{int} = \textbf{f}_\text{ad} + \textbf{f}_\text{sup} = - \frac{1}{\rho} \left< \sum_i \delta_i \nabla_i u(\textbf{r}_1,...,\textbf{r}_N) \right>, \label{eq:fint}
\end{align}
where $\delta_i = \delta(\textbf{r} - \textbf{r}_i) \delta(\boldsymbol{\omega} - \boldsymbol{\omega}_i)$ is used as a shorthand notation with $\delta$ the Dirac delta function, $u$ indicates the interparticle interaction potential, $\nabla_i$ is the derivative with respect to position $\textbf{r}_i$ of the $i=1,...,N$ particle and $\left< \cdot \right>$ is an average in steady state. 
The adiabatic force field $\textbf{f}_\text{ad}$ is defined by the right-hand side of Eq. \eqref{eq:fint} but taken in an equilibrium system  under the influence of an ``adiabatic'' external potential that generates the true density profile $\rho$ \cite{pft2013,fortini2014,renner2019}. Here the corresponding equilibrium system has no flow ($s=0$).
Because of the rotational symmetry of spherical particles $\textbf{f}_\text{ad}$ is independent of the particle orientation $\boldsymbol{\omega}$ for spherical particles as considered here.  
From classical density functional theory \cite{evans1979}, applied to the adiabatic system, it is known that $\textbf{f}_\text{ad}$ is a gradient field obtained as $\textbf{f}_\text{ad} = - \nabla \mu_\text{ad}$ \cite{footnote1}. 
   	
The superadiabatic force field is defined as the difference $\textbf{f}_\text{sup} = \textbf{f}_\text{int} - \textbf{f}_\text{ad}$, cf.\ Eq. \eqref{eq:fint}. From power functional theory \cite{pft2013} follows that $\textbf{f}_\text{sup}$ is a functional of the density profile, but also of the velocity profile.

We split Eq. \eqref{eq:force} into a flow equation and a structural equation, given, respectively, by
\begin{align}
\gamma \textbf{v} = \textbf{f}_\text{flow} + \gamma s \boldsymbol{\omega}, \label{eq:flow} \\
0 = \textbf{f}_\text{id} + \textbf{f}_\text{ad} + \textbf{f}_\text{struc}, \label{eq:struc}
\end{align}
where the superadiabatic force field is the sum of a flow and a structural contribution, $\textbf{f}_\text{sup} = \textbf{f}_\text{flow} + \textbf{f}_\text{struc}$. The splitting is unique. 
The superadiabatic flow force field $\textbf{f}_\text{flow}$ describes the influence of the internal interactions on the flow. 
The structural force field $\textbf{f}_\text{struc}$ is that part of the total internal force field that influences the spatial structure, together with the adiabatic force field $\textbf{f}_\text{ad}$ and the ideal term (which is small in the present situation). Note that it is the functional dependence of $\textbf{f}_\text{sup}$ and hence of $\textbf{f}_\text{struc}$ on velocity which renders Eq. \eqref{eq:struc} (highly) nontrivial. 
Since $\textbf{f}_\text{id}$ and $\textbf{f}_\text{ad}$ are gradient contributions, $\textbf{f}_\text{struc}$ necessarily needs to be a gradient field, $\textbf{f}_\text{struc} = - \nabla \mu_\text{struc}$, which defines $\mu_\text{struc}$ as the negative integral of $\textbf{f}_\text{struc}$. Integrating Eq. \eqref{eq:struc} in space thus leads to 
\begin{align}
\mu_\text{id} + \mu_\text{ad} + \mu_\text{struc} = \mu_b =\text{const}, \label{eq:mub}
\end{align}
where $\mu_b$ is the constant value in the bulk fluid and the sum determines the total chemical potential.
The difference to the equilibrium situation is the dependence of $\mu_\text{struc}$ on the (nonvanishing) flow profile. Conceptually the three chemical potential contributions play the same role as in equilibrium in that their respective gradient is a force field.

The ideal chemical potential $\mu_\text{id} = k_\text{B}T\ln \rho$ is for simplicity reduced to the orientation-independent expression
\begin{align}
\mu_\text{id} =  k_\text{B} T \ln \rho_0 \label{eq:muid}
\end{align}
with the rotational averaged density $\rho_0 = \int \mathrm{d} \boldsymbol{\omega} \; \rho/ 2 \pi$. 
The approximation is reasonable, since the ideal chemical potential is  numerically small in the present situation, as is the corresponding ideal diffusive force (see, e.g.,\ Ref. \cite{bialke2015}). Furthermore $\rho_0$ is a main contribution of the Fourier decomposed density $\rho$ and both densities $\rho$ and $\rho_0$ coincide in bulk.
Since within the used approximations $\mu_\text{id}$ and $\mu_\text{ad}$ are rotationally invariant, Eq. \eqref{eq:mub} implies that $\mu_\text{struc}$ and, hence, $\textbf{f}_\text{struc}$ are also independent of orientation. 

We further discriminate between local and nonlocal contributions in Eq. \eqref{eq:mub}. The ideal chemical potential $\mu_\text{id}$ is a purely local expression and $\mu_\text{ad}$ is also a local term since we base it on a local density approximation. Further nonlocal contributions to $\mu_\text{ad}$ were found to be negligible in the present case. Hence the only considerable nonlocal contribution is contained in $\mu_\text{struc}$, which we split into a sum of local and nonlocal terms, $\mu_\text{struc} = \mu^\text{loc}_\text{struc} + \mu_\text{nloc}$.
The nonlocal superadiabatic chemical potential is approximated as the lowest order gradient contribution,
\begin{align}
\mu_\text{nloc} = - \nabla \cdot \left( m \nabla \rho_0 \right)  + \frac{1}{2} \left(\nabla m \right)\cdot\left(\nabla \rho_0 \right), \label{eq:nloc}
\end{align}
where the coefficient $m$ can depend on density $\rho_0$ and on velocity $\textbf{v}$. Note that $\mu_\text{nloc}$ vanishes in both bulk phases due to the constant density $ \rho_b = \rho_g,\rho_l$, where $\rho_g$ and $\rho_l$ are the constant densities in the gas and liquid phase. Thus bulk chemical potential and local chemical potential coincide in bulk, $\mu_b = \mu_\text{loc}(\rho_b)$.

For the local chemical potential, $\mu_\text{loc} =\mu_\text{id}+\mu_\text{ad}+\mu^\text{loc}_\text{struc}$, the corresponding  nonequilibrium local pressure, $P_\text{loc} = P_\text{id}+P_\text{ad}+P^\text{loc}_\text{struc}$, can be obtained from the Gibbs-Duhem relation \cite{footnote2}
\begin{align}
\frac{\partial P_\text{loc}}{\partial \rho_0} = \rho_0 \frac{\partial \mu_\text{loc}}{\partial \rho_0}. \label{eq:gibbs}
\end{align}

From Eqs. \eqref{eq:mub} and \eqref{eq:nloc} follows directly that $\mu_\text{loc} (\rho_l) = \mu_\text{loc}(\rho_g) = \mu_b$ and using the Gibbs-Duhem relation \eqref{eq:gibbs} leads to $P_\text{loc}(\rho_l) = P_\text{loc}(\rho_g) =P_b$. The combination of both relations allows us to determine both coexistence densities $\rho_g$ and $\rho_l$ and hence the phase diagram of the system, cf.\ Ref. \cite{schmidt2019}.

As we have identified the structural gradient force contributions, we can proceed in a purely mechanical way. Hence the gas-liquid interfacial tension is given by \cite{widom,evans1979}
\begin{align}
\gamma_{\rm gl} = \int \mathrm{d} x \; \left[ \frac{m}{2}\left(\nabla \rho_0 \right)^2 -W \right]. \label{eq:gamma}
\end{align}

Equation \eqref{eq:gamma} consists of a nonlocal and a local part. 
The first, nonlocal contribution results from an (interfacial) square gradient expansion with coefficient $m$. 
The second, local term is given as
\begin{align}
-W = \psi - \psi_b = \left(\mu_\text{loc}-\mu_b \right) \rho_0 - \left(P_\text{loc}-P_b \right), \label{eq:w}
\end{align} 
where $\psi = \mu_\text{loc} \rho_0 - P_\text{loc}$ and $\psi_b = \mu_b \rho_0 - P_b$ contain the above introduced nonequilibrium (local) chemical potential and pressure. Note that $\psi_b$ is not a constant bulk contribution, since $\rho_0$ still depends on $x$. In equilibrium $\psi$ can be identified as the local Helmholtz free-energy density and $\psi_b$ is the corresponding double tangent line.

The chemical potential balance Eq. \eqref{eq:mub} can then be rewritten as
\begin{align}
\frac{\partial W}{\partial \rho_0} + \nabla \cdot \left( m \nabla \rho_0 \right) - \frac{1}{2} \left(\nabla m \right)\cdot\left(\nabla \rho_0\right) = 0, \label{eq:newton}
\end{align}
where we used Eq. \eqref{eq:nloc} to express the nonequilibrium  chemical potential and the derivative of Eq. \eqref{eq:w} with respect to density, $-\partial W/\partial \rho_0 = \mu_\text{loc} -  \mu_b$.
The first integral with respect to $x$ of Eq. \eqref{eq:newton} is 
\begin{align}
W + \frac{1}{2} m \left( \frac{\partial \rho_0}{\partial x} \right)^2 = 0, \label{eq:integral}
\end{align}
where we used the planar symmetry of the density $\rho_0$ to simplify the spatial derivative $\nabla$ to $\hat{\textbf{e}}_x \partial / \partial x$. 
Rewriting the interfacial tension \eqref{eq:gamma} with relation \eqref{eq:integral} leads to three alternative forms:
\begin{align}
\gamma_{\rm gl} &= \int\limits_{-\infty}^{\infty} m \, \left( \frac{\partial \rho_0}{\partial x} \right)^2 \; \mathrm{d} x  \label{eq:y1}\\
&= -2 \int\limits_{- \infty}^{\infty} W \; \mathrm{d}x  \label{eq:y2}\\
&= \int\limits_{\rho_g}^{\rho_l} \sqrt{-2 \, m \, W} \; \mathrm{d}\rho_0.  \label{eq:y3}
\end{align}
The numerical values of Eqs. \eqref{eq:y1}-\eqref{eq:y3} only coincide if the functions $\rho_0$, $m$, and $W$ are chosen reasonably and satisfy Eq.\ \eqref{eq:integral}.
Thus whether a choice of these three functions is appropriate can
be gauged by the agreement of the value for $\gamma_{\rm gl}$
obtained from either of Eqs. \eqref{eq:y1}-\eqref{eq:y3}. This provides
a check for the approximations for $m$ and $W$ as introduced below.

Equation \eqref{eq:y1} does not depend on the local contribution $W$ and is thus referred to as the nonlocal route. The relation \eqref{eq:y2} is independent of the coefficient $m$ of the nonlocal term. It is denoted as the local route, as the integrand is the local quantity $W$. Expression \eqref{eq:y3} is called the no-profile route, as it is independent of the density distribution $\rho_0$. In practice it can be useful to calculate $\gamma_{\rm gl}$ without knowledge of $\rho_0$.
In the equilibrium limit of passive particles ($s = 0$) and vanishing particle velocity $\textbf{v} = 0$, our expressions for the interfacial tension coincide with the known equilibrium relations, cf.,\ e.g., Ref. \cite{widom}. 

We apply our general theory for the nonequilibrium interfacial tension to a system of two-dimensional active particles which interact via a Weeks-Chandler-Anderson potential. This is a Lennard-Jones potential cut at its minimum and shifted to be continuous. The corresponding energy scale is $\epsilon$ and the characteristic length scale $\sigma$ is also referred to as the diameter of the spherical particles.  
The orientational motion is freely diffusive with rotational
diffusion constant $k_BT/\gamma^{\boldsymbol{\omega}}$, where
$\gamma^{\boldsymbol{\omega}}$ denotes the rotational friction
constant. 
The rotational averaged density can be approximated with high accuracy as a hyperbolic tangent profile \cite{utrecht2018}
\begin{align}
\rho_0(x) = \frac{\rho_g+\rho_l}{2}+\frac{\rho_l -\rho_g}{2} \tanh \left(\frac{x}{\lambda}\right),
\end{align}
where $\lambda$ indicates the interfacial width. The coexistence densities $\rho_g$ and $\rho_l$ were determined from the pressure and chemical potential balance at theoretical coexistence and coincide with results from simulations very well \cite{schmidt2019,schmidt2019footnote}.
Note that we do not consider the flow terms \eqref{eq:flow} to the force balance here as we focus on the interfacial tension. 

The chemical potential contributions are chosen in accordance with Ref. \cite{schmidt2019}.
The ideal term is given by relation \eqref{eq:muid}.
For the adiabatic chemical potential we use the local density approximation on a scaled particle theory for two-dimensional hard disks \cite{barker}. This yields 
\begin{align}
\mu_{\text{ad}} = k_\text{B} T \left[ -\ln(1-\eta')+\eta' (3-2\eta')/(1-\eta')^2 \right], \label{eq:muad}
\end{align}
where the rescaled packing fraction $\eta' = 0.8 \eta$ models the soft Weeks-Chandler-Anderson potential. The packing fraction $\eta = \rho_0 / \rho_\text{j}$ and $\rho_\text{j}=\text{const}$ indicates the jamming density, where the motion comes to arrest.
The remaining $\mu_\text{struc}$ corresponds to the quiet life chemical potential \cite{schmidt2019}, which in homogeneous bulk is given as 
\begin{align}
\mu_\text{struc}^b = \frac{e_1 \gamma \gamma^{\boldsymbol{\omega}}}{2 k_\text{B}T} v_b^2 \frac{\rho_b}{ \rho_{\rm j}}, \label{eq:nu3b}
\end{align}
where the strength is determined by the dimensionless constant $e_1$. The expression \eqref{eq:nu3b} is linear in bulk density $\rho_b$, quadratic in the bulk speed $v_b$ and the resulting force acts toward the liquid phase. Note that due to its velocity dependence $\mu_\text{struc}^b$ is a genuine nonequilibrium expression. 
To obtain the local structural chemical potential we expand Eq. \eqref{eq:nu3b} across the interface using the orientational averaged density $\rho_0$ instead of $\rho_b$ and the known linear decrease $v_\text{loc} = s(1-\rho_0/\rho_\text{j})$ \cite{speck2015} for the speed $v_b$. This yields 
\begin{align}
  \mu_\text{struc}^{\text{loc}}
  = \frac{e_1}{6}  \text{Pe}^2 k_\text{B}T \left( 1 - \frac{\rho_0}{\rho_\text{j}} \right)^2 \frac{\rho_0}{ \rho_{\rm j}},  \label{eq:nu3loc}
\end{align}
where the introduced Péclet number is $\text{Pe}= s \sigma \gamma/k_\text{B}T =3 s \gamma^{\boldsymbol{\omega}}/k_\text{B}T \sigma$. This dimensionless constant relates active swimming to rotational diffusion.

\begin{figure}[bt]
\includegraphics[width=0.47\textwidth]{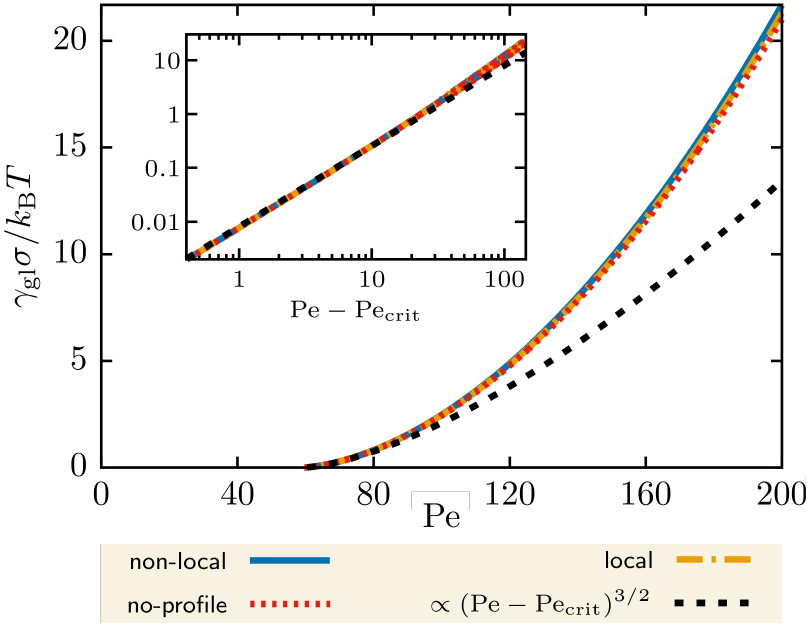}
\vspace*{-2.5mm}
\caption{\label{fig:tension} Interfacial tension $\gamma_{\rm gl}$ determined from the nonlocal route Eq. \eqref{eq:y1} (full blue line), from the local route Eq. \eqref{eq:y2} (dash-dotted yellow line), and from no-profile route Eq. \eqref{eq:y3} (dotted red line) in dependence of the Péclet number Pe. Close to the critical point the tension increases with a critical exponent of $3/2$, as indicated by the dashed black line. The inset also shows $\gamma_{\rm gl}$ but in a double-logarithmic plot and the $x$ axis is shifted by the critical Péclet number and, hence, is $\text{Pe}-\text{Pe}_\text{crit}$.}
\end{figure}

The nonlocal chemical potential is approximated in the simplest way, with a constant coefficient $m = e_2  \text{Pe}^2 k_\text{B}T / 6 \rho_\text{j}^2$, such that Eq. \eqref{eq:nloc} simplifies to $\mu_\text{nloc} =  - m \nabla^2 \rho_0$ and one obtains
\begin{align}
  \mu_\text{nloc}= -\frac{e_2 \text{Pe}^2}{6}\frac{k_\text{B}T}{\rho_{\rm j}} \nabla^2 \frac{\rho_0}{\rho_{\rm j}},  \label{eq:nu3nloc}
\end{align}
where the amplitude is determined by the dimensionless constant $e_2$.
One can show within the power functional framework \cite{pft2013, krinninger2019}, that $\mu_\text{struc}$ is an intrinsic quantity and can be written as a kinematic functional, hence only dependent on density $\rho$ and velocity $\textbf{v}$. Therefore, $\mu_\text{struc}^\text{loc}$ and $\mu_\text{nloc}$ are ``naturally'' independent of the swim speed and Eq. \eqref{eq:nu3loc} can be expressed without $s$ as an intrinsic expression \cite{schmidt2019,footnote3}.
The local pressure can be determined straightforwardly from the Gibbs-Duhem relation Eq. \eqref{eq:gibbs}.

The parameters of the system are chosen as follows. The system is at temperature $k_\text{B}T/\epsilon=0.5$, has a rotational friction coefficient $\gamma^{\boldsymbol{\omega}}/ \gamma \sigma^2 = 1/3$, a jamming density of $\rho_\text{j} 2 \pi \sigma^2 = 1.4$ and the dimensionless prefactors $e_1 = 0.0865$ and $e_2=0.0385$.
Requiring $e_2$ to be constant and the chemical potential balance \eqref{eq:mub} to be satisfied, the interfacial width $\lambda$ is determined.
The swim speed $s$ changes with Péclet number, $\text{Pe}=s \sigma \gamma / k_\text{B} T$, while the other parameters are kept constant.
We use the approximations for the orientational averaged density profile $\rho_0$, the chemical potential contributions Eqs. \eqref{eq:muid}, \eqref{eq:muad}, \eqref{eq:nu3loc} and \eqref{eq:nu3nloc} and the corresponding pressures to determine the interfacial tension by evaluating the expressions \eqref{eq:y1}-\eqref{eq:y3}.

The results from the three methods are displayed in Fig. \ref{fig:tension}.
We find the behavior of the function $W$ and of the interfacial tension to be qualitatively similar to what is found in equilibrium \cite{widom}. Figure \ref{fig:tension} shows $\gamma_{\rm gl}$ as a function of $\text{Pe}$. The interfacial tension is only different from zero for Péclet numbers larger than $\text{Pe}_\text{crit} = 59.3$ \cite{schmidt2019}, when the system phase separates.
Here the critical value of the P\'eclet number \cite{schmidt2019}
  is determined by the magnitude of $e_1$. 
The tension increases with rising particle activity and hence with the Péclet number (cf. Fig. \ref{fig:tension}). Close to the critical point $\gamma_{\rm gl}$ increases with a critical exponent of 3/2, as indicated by the black dashed line. This corresponds to the theoretical mean-field coefficient of the van der Waals theory, which might be expected since there are many similarities between both descriptions. In order to emphasize the agreement of the interfacial tension with a function proportional to $\left(\text{Pe}-\text{Pe}_\text{crit}\right)^{3/2}$, both quantities are displayed in a double logarithmic plot (cf. the inset of Fig. \ref{fig:tension}). For Péclet numbers close to the critical point, the functions nearly have the same slope.
Far from the critical point the interfacial tension increases faster than with the critical exponent. 
For a detailed simulation study of the bulk critical behavior of active Brownian particles, see Ref. \cite{speck2018}. 

The values of the tension are positive, $\gamma_{\rm gl}>0$, which directly explains the stability of the interface. 
This is in contrast to Bialké et al. \cite{bialke2015}, who calculated a negative interfacial tension using the pressure tensor. 
The results for three different methods, the nonlocal route Eq. \eqref{eq:y1}, the local route Eq. \eqref{eq:y2}, and the no-profile route Eq. \eqref{eq:y3}, agree to a very satisfying degree (cf.\ Fig. \ref{fig:tension}). Even far from equilibrium, for example, at $\text{Pe}=200$, the respective results deviate by only about $3\%$. This indicates that the chemical potential balance \eqref{eq:mub} and hence the structural force balance $\eqref{eq:struc}$ are both satisfied with very good accuracy.
Finally, the splitting \eqref{eq:flow} and \eqref{eq:struc} [together with \eqref{eq:gamma} within a square gradient approximation] forms a general route toward the interfacial tension of out-of-equilibrium interfaces.
We have also ascertained that the ``flow'' equation of motion
  \eqref{eq:flow} creates a vanishing contribution to the interfacial
  tension in the present system since after orientational integration the associated pressure
   tensor contributions either vanish or are isotropic.
Hence the splitting \eqref{eq:flow} and \eqref{eq:struc} does not
imply omission of any relevant terms.

Because of the square gradient character of our treatment, we do not find layering effects at the interface, which requires us to take account of nonlocal interfacial packing effects \cite{tarazona2001}.
Furthermore, our treatment yields the ``intrinsic density profile'' \cite{tarazona2005, tarazona2004}, as large scale capillary wave fluctuations are neglected.
Thus, interesting future work could be devoted to studying capillary wave fluctuations and the wave vector dependence of the interfacial tension \cite{tarazona2005, tarazona2004}.
Furthermore, it would be interesting to relate our treatment
to that presented in Ref.~\cite{solon2018} and to consider fluctuations beyond mean field that could alter
the value of the critical scaling exponent.


\begin{thebibliography}{31}

\bibitem{vrij1997} A. Vrij, Physica (Amsterdam) \textbf{235A}, 120 (1997).
%Demixed phases of colloid plus polymer systems in a common solvent Calculation of the interfacial tension

\bibitem{brader2000} J. M. Brader and R. Evans, Europhys. Lett. \textbf{49}, 678 (2000).
%The fluid-fluid interface of a model colloid-polymer mixture

\bibitem{brader2002} J. M. Brader, R. Evans, M. Schmidt, and H. L{\"o}wen, J. Phys.: Condens. Matter \textbf{14}, L1 (2002).
%Entropic wetting and the fluid-fluid interface of a model colloid-polymer mixture

\bibitem{hoog1999} E. H. A. de Hoog and H. N. W. Lekkerkerker, J. Phys. Chem. B \textbf{103}, 5274 (1999).
%Measurement of the Interfacial Tension of a Phase-Separated Colloid− Polymer Suspension

\bibitem{hoog2001} E. H. A. de Hoog and H. N. W. Lekkerkerker, J. Phys. Chem. B \textbf{105}, 11636 (2001).
%Breakup of an elongated droplet in a centrifugal field

\bibitem{aarts2004} D. G. A. L. Aarts, M. Schmidt, and H. N. W. Lekkerkerker, Science \textbf{304}, 847 (2004).
%Direct visual observation of thermal capillary waves

\bibitem{aarts2008} D. G. A. L. Aarts and H. N. W. Lekkerkerker, J. Fluid Mech. \textbf{606}, 275 (2008).
%Droplet coalescence: drainage, film rupture and neck growth in ultralow interfacial tension systems

\bibitem{setu2013} S. A. Setu, I. Zacharoudiou,  G. J. Davies,  D. Bartolo,  S. Moulinet,  A. A. Louis,  J. M. Yeomans, and D. G. A. L. Aarts, Soft Matter \textbf{9}, 10599 (2013).
%Viscous fingering at ultralow interfacial tension

\bibitem{farage2015} T. F. F. Farage, P. Krinninger, and J. M. Brader, Phys. Rev. E \textbf{91}, 042310 (2015).
%Effective interactions in active Brownian suspensions

\bibitem{brader2017} R. Wittmann, U. M. B. Marconi, C. Maggi, and J. M. Brader, J. Stat. Mech. (2017) 113208. 
%Effective equilibrium states in the colored-noise model for active matter II. A unified framework for phase equilibria, structure and mechanical properties 
  
\bibitem{speck2015} T. Speck, A. M. Menzel, J. Bialk{\'e}, and H. L{\"o}wen, J. Chem. Phys., \textbf{142}, 224109 (2015).
%Dynamical mean-field theory and weakly non-linear analysis for the phase separation of active Brownian particles

\bibitem{utrecht2018} S. Paliwal, J. Rodenburg, R. van Roij, and M. Dijkstra, New J. Phys. \textbf{20}, 015003 (2018).
%Chemical potential in active systems: predicting phase equilibrium from bulk equations of state?

\bibitem{buttinoni2013} I. Buttinoni, J. Bialk{\'e}, F. K{\"u}mmel, H. L{\"o}wen, C. Bechinger, and T. Speck, Phys. Rev. Lett. \textbf{110}, 238301 (2013).
%Dynamical Clustering and Phase Separation in Suspensions of Self-Propelled Colloidal Particles

\bibitem{parry2016} C. Rascon, A. O. Parry, and D. G. A. L. Aarts, 	Proc. Natl. Acad. Sci. U.S.A. \textbf{113}, 12633 (2016).
%Geometry-induced capillary emptying

\bibitem{parry2016cwt} A. O. Parry, C. Rascon, and R. Evans, J. Phys.: Condens. Matter \textbf{28}, 244013 (2016).
%The local structure factor near an interface; beyond extended capillary-wave models.

\bibitem{parry2019} A. O. Parry and C. Rascon, Nat. Phys. \textbf{15}, 287 (2018).
%The Goldstone mode and resonances in the fluid interfacial region

\bibitem{oettel2010} B. J. Block, S. K. Das, M. Oettel, P. Virnau, and K. Binder, J. Chem. Phys. \textbf{133}, 154702 (2010).
%Curvature dependence of surface free energy of liquid drops and bubbles: A simulation study

\bibitem{oettel2007} F. Bresme and M. Oettel, J. Phys. Condens. Matter \textbf{19}, 413101 (2007).
%Nanoparticles at fluid interfaces

\bibitem{roth2004} P. M. K{\"o}nig, R. Roth, and K. R. Mecke, Phys. Rev. Lett. \textbf{93}, 160601 (2004).
%Morphological thermodynamics of fluids: Shape dependence of free energies

\bibitem{tarazona2001} E. Chac{\'o}n, M. Reinaldo-Falag{\'a}n, E. Velasco, P. Tarazona, Phys. Rev. Lett. \textbf{87}, 166101 (2001).
%Layering at free liquid surfaces

\bibitem{tarazona2004} P. Tarazona and E. Chac{\'o}n, Phys. Rev. B \textbf{70}, 235407 (2004).
% Monte Carlo intrinsic surfaces and density profiles for liquid surfaces

\bibitem{tarazona2005} E. Chac{\'o}n and P. Tarazona, J. Phys. Condens. Matter \textbf{17}, S3493 (2005).
%Characterization of the intrinsic density profiles for liquid surfaces

\bibitem{oettel2012} A. Tr{\"o}ster, M. Oettel, B. Block, P. Virnau, and K. Binder, J. Chem. Phys. \textbf{136}, 064709 (2012).
%Numerical approaches to determine the interface tension of curved interfaces from free energy calculations

\bibitem{bialke2015} J. Bialk{\'e}, J. T. Siebert,
H. L{\"o}wen, and T. Speck, Phys. Rev. Lett. \textbf{115}, 098301 (2015).
%Negative Interfacial Tension in Phase-Separated Active Brownian Particles

\bibitem{patch2018} A. Patch, D. M. Sussman, D. Yllanes, and M. C. Marchetti, Soft Matter {\bf 14}, 7435 (2018).

\bibitem{lee2017} C. F. Lee, Soft Matter {\bf 13}, 376 (2017).

\bibitem{solon2018} A. P. Solon, J. Stenhammar, M. E. Cates, Y. Kafri, and J. Tailleur, New J. Phys. {\bf 20}, 075001  (2018).

\bibitem{marconi2015} U. M. B. Marconi and C. Maggi, Soft Matter {\bf 11}, 8768 (2015).

\bibitem{marconi2016} U. M. B. Marconi, C. Maggi, and S. Melchionna, Soft Matter {\bf 12}, 5727 (2016).

\bibitem{das2019} S. Das, G. Gompper, and R. G. Winkler, Sci. Rep. {\bf 9}, 6608 (2019).

\bibitem{paliwal2017acvtiveLJinterface} S. Paliwal, V. Prymidis, L. Filion, and M. Dijkstra, J. Chem. Phys. {\bf 147}, 084902 (2017).

\bibitem{widom} J.~S. Rowlinson and B. Widom, {\it Molecular Theory of Capillarity }  (Dover, New York, 2002).
%Molecular Theory of Capillarity

\bibitem{krinninger2019} P. Krinninger and M. Schmidt, J. Chem. Phys \textbf{150}, 074112 (2019).
%Power functional theory for active Brownian particles: general formulation and power sum rules

\bibitem{krinninger2016} P. Krinninger, M. Schmidt, and J. M. Brader, Phys. Rev. Lett. \textbf{117}, 208003 (2016).
%Nonequilibrium Phase Behavior from Minimization of Free Power Dissipation

\bibitem{pft2013} M. Schmidt and J. M. Brader, J. Chem. Phys. \textbf{138}, 214101 (2013).
%Power functional theory for Brownian dynamics

\bibitem{fortini2014} A. Fortini, D. de las Heras, J. M. Brader, and M. Schmidt, Phys. Rev. Lett. \textbf{113}, 167801 (2014).
%Superadiabatic Forces in Brownian Many-Body Dynamics

\bibitem{renner2019} D. de las Heras, J. Renner, and M. Schmidt, Phys. Rev. E \textbf{99}, 023306 (2019).
%Custom flow in overdamped Brownian Dynamics

\bibitem{evans1979} R. Evans, Adv. Phys. \textbf{28}, 143 (1979).
%The nature of the liquid-vapour interface and other topics in the statistical mechanics of non-uniform, classical fluids
 
\bibitem{footnote1}
Within density functional theory, using the Helmholtz excess free-energy functional $F_\text{exc}[\rho]$, the internal chemical potential $\mu_\text{ad} = \delta F_\text{exc}[\rho] / \delta \rho$ can be written as a functional of (only) the density profile and independent of external forces.

\bibitem{footnote2}
The Gibbs-Duhem equation results from identifying the negative gradient of a chemical potential $\mu$ as a force and the negative gradient of a pressure $P$ as a force density. The combination of both relations leads to $- \rho \nabla \mu = - \nabla P$, where the spatial derivative can be rewritten as $\rho \partial \mu/ \partial \rho_0 \nabla \rho_0 = \partial P/ \partial \rho_0 \nabla \rho_0$. Simplification and averaging over orientation gains Eq. \eqref{eq:gibbs} in case of local and rotational independent $\mu$ and $P$.

\bibitem{schmidt2019} S. Hermann, P. Krinninger, D. de las Heras, and M. Schmidt, Phys. Rev. E {\bf 100}, 052604 (2019).
%Phase coexistence of active Brownian particles: Anything for a quiet life

\bibitem{schmidt2019footnote} The simulation results presented in Ref.~\cite{schmidt2019} are based on $N=2000$ particles in rectangular boxes of varying aspect ratio 2.5, 5, and 10. The systems were initialized in configurations with the interface running along the short direction of the simulation box.

\bibitem{barker} J. A. Barker and D. Henderson, Rev. Mod. Phys. \textbf{48}, 587 (1976).
%What is a "liquid"? Understanding the states of matter

\bibitem{footnote3} In order to express the swim speed with internal quantities, one can use the forward speed $v_\text{f} = \int \mathrm{d} \boldsymbol{\omega} \textbf{J} \cdot \boldsymbol{\omega} / 2 \pi \rho_0$, the orientational integrated projection of the translational  current $\textbf{J}$ on the particle orientation $\boldsymbol{\omega}$. It is approximately given as $v_\text{f} = s(1-\rho_0/\rho_\text{j})/[1+\xi (\nabla \rho_0)^2\rho_0/\rho_\text{j}]$ \cite{schmidt2019}, where the constant $\xi>0$ sets the amplitude of a square gradient expansion term. Hence the linear decrease in speed $v_\text{loc} = s(1-\rho_0/\rho_\text{j})$ in \eqref{eq:nu3b} can be replaced by the intrinsic expression $v_\text{f}[1+\xi (\nabla \rho_0)^2\rho_0/\rho_\text{j}]$.

\bibitem{speck2018} J. T. Siebert, F. Dittrich, F. Schmid, K. Binder, T. Speck, and P. Virnau, Phys. Rev. E \textbf{98}, 030601(R) (2018).
%Critical behavior of active Brownian particles

\end{thebibliography}
\end{document}